\documentclass[12pt,aps]{revtex4}
\usepackage{amssymb}
\usepackage{amsmath}
\usepackage{graphicx}
\usepackage{epsfig,amsmath}
\usepackage[bookmarksnumbered,linktocpage,pdfstartview=FitH]{hyperref}

\begin{document}

\title{Statistical signature of vortex filaments in classic turbulence: dog
or tail?}
\author{Sergey K. Nemirovskii\thanks{%
email address: nemir@itp.nsc.ru}}
\affiliation{Institute of Thermophysics, Lavrentyev ave., 1, 630090, Novosibirsk, Russia\\
and Novosibirsk State University, Pirogova Str., 2, 630090, Novosibirsk,
Russia}
\date{\today }

\begin{abstract}
The title of this paper echoes the title of a paragraph in the famous book
by Frisch on classical turbulence. In the relevant chapter, the author
discusses the role of the statistical dynamics of vortex filaments in the
fascinating problem of turbulence and the possibility of a breakthrough in
constructing an advanced theory. This aspect arose due to the large amount
of evidence, both experimental and numerical, that the vorticity field in
turbulent flows has a pronounced filamentary structure. In fact, there is
unquestionably a strong relationship between the dynamics of chaotic vortex
filaments and turbulent phenomena. However, the question arises as to
whether the basic properties of turbulence (cascade, scaling laws. etc.) are
a consequence of the dynamics of the vortex filaments (the `dog' concept),
or whether the latter have only a marginal significance (the `tail'
concept). Based on well-established results regarding the dynamics of
quantized vortex filaments in superfluids, we illustrate how these dynamics
can lead to the main elements of the theory of turbulence. We cover key
topics such as the exchange of energy between different scales, the possible
origin of Kolmogorov-type spectra and the free decay behavior.%
\end{abstract}

\maketitle

\section{Introduction}

\subsection{Quantum turbulence vs classical turbulence}

The idea that classical turbulence (CT) can be modeled by the dynamics of a
set of slim vortex tubes (or vortex sheets) has been discussed for quite a
long time (a good exposition of the prehistory of that question can be found
in the famous books by Frisch \cite{Frisch1995}, and by Chorin \cite%
{Chorin1994} ). The main motivation for this idea is related to incredible
complexity in the traditional formulation of this tantalizing problem.
Indeed, as Migdal \cite{Migdal1986} wrote: "Hydrodynamics over the centuries
has been described by partial differential equations. This description
turned out to be adequate for laminar flows, but faced difficulties in the
phenomena of turbulence. The source of these difficulties is the choice of
velocity field components $\mathbf{v(r},t)$ as dynamic variables. This field
behaves stochastically in turbulent flows, which makes the differential
equations useless. The required number of degrees of freedom exceeds the
capabilities of any computer. The vortex filaments can be considered as
elementary excitations of the turbulent flow. Their dynamics, although
unusual, is actually more simple than the dynamics of waves and particles,
and requires less computer resources".

Beside this motivation regarding new breakthrough methods in the theory of
CT, there are also numerous observations that the vorticity field in
turbulent flows has a pronounced filamentary structure. The range of
evidences extends from 500-year-old drawings of eddy motion made by Leonardo
da Vinci (see e.g., \cite{Frisch1995} or \cite{Tsubota2009a}) up to modern
powerful numerical simulations (see, e.g. \cite{Vincent1991}). In fact,
there is unquestionably a relationship between the dynamics of a chaotic
vortex filament and turbulent phenomena. However, the question arises as to
whether the properties of turbulence are a consequence of the dynamics of
the vortex filaments, or whether the latter have only a marginal
significance. To highlight this issue, Frisch \cite{Frisch1995} entitled a
chapter in his book devoted to this topic as "Statistical signature of
vortex filaments: Dog or tail?"

In classical fluids, the concept of thin vortex tubes is rather the fruitful
mathematical model. Quantum fluids, where vortex filaments are real objects,
provide an excellent opportunity for studying the question of whether the
chaotic dynamics of a set of quantized vortex lines (quantum turbulence or
QT) can reproduce the properties of real hydrodynamic turbulence. This
feature of QT is usually referred as the quasi-classical regime.

In principle, there is large huge amount of scientific researches on quantum
turbulence and its relation to classic turbulence (CT) in conventional
fluids, the according material is summarized in series of recent review
articles \cite{Vinen2000}, \cite{Kobayashi2005},\cite{Vinen2010},\cite%
{Skrbek2012},\cite{Tsubota2013},\cite{Barenghi2014a},\cite{Walmsley2007}.
However, the vast majority of research in this area uses the ideas and
results of CT to explain phenomena in QT, i.e. CT $\rightarrow $ QT. It can
be said that the relationship between QT and CT is such that the former is a
recipient and the latter is a donor. In the current article the, opposite
direction is chosen, it is we who are trying to state what properties of QT
phenomena can be used to explain some features of CT, i.e. QT $\rightarrow $
CT. This statement corresponds to the question submitted in the manuscript
title -- DOG OR TAIL.

\subsection{What this article is and is not about}

The main goal of this paper is to discuss which elements of the quantized
vortex dynamics, established and studied recently, would lead to the main
ingredients of theory of classic turbulence. Before proceeding with the
content of the work, it seems useful to make some appropriate reservations,
based on preliminary discussions.

\qquad First of all this work does not claim that the chaotic dynamics of
discrete (quantized) vortex filaments can simulate classical turbulence.
Indeed, as it was written above this idea had been discussed by other
scientists, who observed, or supposed, or obtained from numerical
experiments, that the vorticity field has a filamentary structure. I pursue
a more modest goal, which is simply to disclose the \textquotedblleft dog's
concept\textquotedblright\ on the basis of well-known facts from the
researches of discrete quantum vortices in superfluids. And this is not a
review article, rather it is application of the well established results on
dynamics of quantized vortices to concrete physical problem.

\qquad Secondly, in order to clarify the question formulated by Frisch \cite%
{Frisch1995} concerning only vortex filaments, I did not consider the
possible role of other vortex structures, such as vortex sheets or vortex
bundles.

Furthermore, in order to concentrate on the main goal (without distracting
the readers from the basic ideas), I avoid a detailed description of
complicated mathematical calculations. Instead, I refer to the original
papers in which the detailed technique is presented.

\textsl{\ }In addition, since our primary purpose\textsl{\ }was to
understand the essence of turbulent phenomena in terms of vortex filaments,
I confined myself to analytical studies that have quantitative conclusions.
Indeed, although numerical modeling also leads to results that are relevant
to the quasiclassical QT regime (for example, the Kolmogorov spectra), the
origin of these phenomena is hidden behind a numerical procedure.

Finally, I also confine myself to the most pronounced features of the
turbulent flow, such as the exchange of energy between different scales
(Sec. II), Kolmogorov's energy spectra (Section III), and the free decay of
QT (Section IV).

\section{Exchange of energy between different scales.}

\subsection{Kinetics of vortex loops}

The first topic that we will discuss concerns a basic feature of turbulent
flows, i.e. the exchange of energy between different scales. In the case of
QT this effect is quite expected, since the dynamics of quantized vortex
filaments is essentially nonlinear. However, the concrete implementation of
this phenomenon in terms of lines is not clear. In this Section, we will
consider one of the possible mechanisms that leads to the energy cascade,
this is the kinetics of colliding vortex loop, constituting the QT.

This model assumes that the vortex tangle consists of a set of many closed
vortex loops, which undergo an enormous number of reconnections and
self-reconnections, involving (for typical experiments) several millions of
collisions per second (per cm$^{3}$). Thus, in the full statement of the
problem, we are dealing with a set of objects (vortex loops) which do not
have a fixed number of elements and which can be born and die. Additionally
the objects (vortex loops) themselves possess an infinite number of degrees
of freedom with very involved dynamics. Clearly, this problem can hardly be
resolved in the near future, and substantial simplifications are required.
One of the possible treatments of this problem is to impose a condition
whereby the vortex loops have a pre-defined structure, namely the random
walking structure. The idea that one-dimensional topological defects have a
random walking structure is widely used (see, e.g., books by Kleinert \cite%
{Kleinert1990}). For vortex loops in superfluid helium this idea is realized
in form of the so-called generalized Wiener distribution, which takes into
account the anisotropy, polarization and finite curvature \cite%
{Nemirovskii1998}.

\begin{figure}[tbp]
\includegraphics[width=10cm]{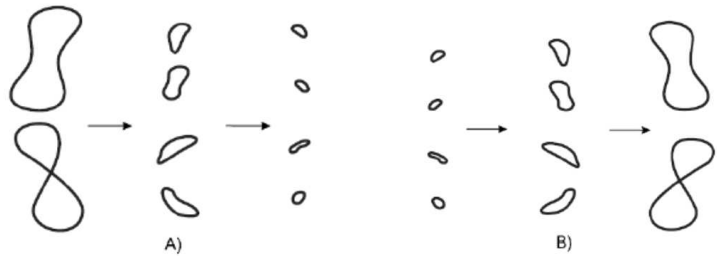}
\caption{ The cascade -like breaking down and merging of Gaussian vortex
loops due to reconnection and self-reconnection processes. }
\label{recombination}
\end{figure}

It is assumed here that the \textquotedblleft own\textquotedblright\
dynamics of the vortex loop are omitted; to some extent, this is absorbed by
the random walk model and the parameters of the generalized Wiener
distribution. We can then consider the evolution of the vortex tangle as the
free motion of vortex loops that have a predetermined random walk structure.
All interactions are reduced to collisions (self-collisions) and
recombinations of loops, as shown in Fig.  \ref{recombination}. This
approach allows for a detailed investigation of the kinetics of vortex
tangle, corresponding studies have been  performed in  \cite%
{Nemirovskii2006,Nemirovskii2008,Nemirovskii2013}.

To demonstrate the existence of energy cascade in space of scales, the usual
variant of the Wiener distribution had been chosen with an\ elementary step $%
\xi _{0}$ of the order of the intervortex distance $\xi _{0}\sim \delta =$ $%
\mathcal{L}^{-1/2}$, where $\mathcal{L}$ is the vortex line density (VLD).

Then, the only degree of freedom which exists is the length $l$ of the loop,
and it is natural to introduce the distribution function of of time
dependent number density of loops $n(l,t)$\ (per unit volume) with length$\ l
$. The temporal evolution of the quantity $n(l,t)$\ can be studied on the
basis of the Boltzmann-type \textquotedblleft balance
equation\textquotedblright , which is a highly nonlinear integral equation.\
Using a special and elegant technique developed by Zakharov (see, e.g. \cite%
{Zakharov1992}), it was demonstrated that the stationary balance equation\
has an exact power-like solution for the distribution function $n(l)$,
namely $n(l)=const\ast l^{-5/2}$\ \cite%
{Nemirovskii2006,Nemirovskii2008,Nemirovskii2013}. This  is not equilibrium
solution; it describes the flux of the length of loops in space of their
sizes. The term "flux" \ here means just the redistribution of length among
the loops due to reconnections. The stated approach generates a number of
predictions that can be associated with the quasiclassical behavior of
quantum turbulence.

The main prediction related to the topic of this paper concerns the constant
flux of energy in the space of scales\textbf{. }The solution described above
corresponds to a non-equilibrium state, and describes a flux $P^{L}(l)$ of
length density $L(l,t)=ln(l,t)$, which represents length accumulated in
loops of size $l$ in $l-$space. Furthermore we assume that due to the
extremely small core radius of the vortex filament the length of vortex line
does not change during\textsl{\ }reconnection event. Conservation of the
vortex line density can be expressed in the form of a continuity equation
for the length density $L(l,t)$

\begin{equation}
\frac{\partial L(l,t)}{\partial t}+\frac{\partial P^{L}(l)}{\partial l}=0.
\label{continuity equation}
\end{equation}

In the stationary case, the expression for the $l$-independent flux $P^{L}$\
can be directly derived from the balance equation, again with the use of the
Zakharov technique \cite{Zakharov1992}. The according calculations were
performed in the papers cited above  \cite%
{Nemirovskii2006,Nemirovskii2008,Nemirovskii2013} and suggest the following
result for the flux of energy $P^{E}$

\begin{equation}
P^{E}=C_{F}\kappa (\kappa \mathcal{L})^{2}.  \label{diss rate}
\end{equation}%
Constant $C_{F}$\ was named in honor of Richard Feynman, who was the first
to discuss the decay of superfluid turbulence due to the cascade-like
breaking down of vortex loops \cite{Feynman1955}. We changed \emph{the} $l$
independent \emph{flux }$P^{L}$\emph{\ \ of the length of loops by the
energy flux }$P^{E}$\emph{, since }in the local induction approximation
(LIA)\ the energy $E$ of a line (up to factor $\rho _{s}\kappa $, where $%
\rho _{s}$ is the superfluid density and $\kappa $ is quantum of circulation
of vortex filament) is equal to its length $L$. Therefore Eq. (\ref%
{continuity equation}) describes the constant flux of the vortex energy $%
P^{E}(l)=const$ in the space of scales (which is here just the space loop
sizes $l$).

The flux $P^{E}$\ consists of two contributions: the first, positive one is
related to the merging of loops, and is responsible for delivering the
energy to large scales, while the second, negative contribution appears to
be due to the breaking-down of loops (see Fig. \ref{recombination}), and
describes a flux of energy to small scales. Depending on the temperature,
either one can prevail; this depends just on the temperature behavior of
structure constants of the vortex tangle. In particular, at very low
temperature $C_{F}\ \approx -0.25$\ (see \cite%
{Nemirovskii2006,Nemirovskii2008,Nemirovskii2013}), the energy flux is
negative, i.e. $P^{E}<0$, and the energy is transferred into the region of
small scales. This corresponds to the direct cascade in classical
turbulence.

A relation of type (\ref{diss rate}) is also known as a Vinen equation. It
describes attenuation of the vortex line density after switching off the
counterflow of He II. It was obtained in a purely phenomenological way using
experimental data. For low temperatures, $C_{F}\ $\ is of the order of $0.1$%
,\ which agrees by the order of the quantity with our estimation.

Each vortex loop induces a three-dimensional velocity field. The whole
ensemble of loops generates a random velocity field that can be considered
turbulent motion. It is clear that large vortex loops create a
three-dimensional velocity field that can be associated with large-scale
motion. Thus, it can be argued that the process of the cascade-like breaking
down of vortex loops can be associated with the main feature of turbulence
-- the constant Kolmogorov-Richardson flux of energy in the space of scales
with the consequent dissipation of kinetic energy at very small scales. As
Feynman wrote "a vortex ring can break up into smaller and smaller rings.
The eventual small rings may be identical to rotons. Then all the energy of
the vortex will eventually end by forming large numbers of rotons, that is,
heat" (see \cite{Feynman1955}, citation is slightly adapted)

\subsection{\emph{\ } Stochastic deformation of loops}

Another mechanism for the exchange of energy between scales, inherent for
turbulent phenomena, is related to the nonlinear dynamics of a single vortex
filament (neglecting the reconnections). We can describe the solution to the
problem of nonlinear chaotic distortions of a vortex loop \cite%
{Nemirovskii2001b} as follows. The chaotic motion of a quantized vortex
filament obeys a Langevin type equation
\begin{equation}
{\frac{{\partial ~}\mathbf{s}{(\xi ,t)}}{{\partial ~t}}}~={\frac{\kappa }{{%
4\pi }}}\int {\frac{[\mathbf{s}(\xi ^{\prime },t)-\mathbf{s}(\xi ,t)]\times
\mathbf{s}_{\xi ^{\prime }}^{\prime }}{{|\mathbf{s}(\xi ^{\prime },t)-%
\mathbf{s}(\xi ,t)|^{3}}}}d\xi ^{\prime }+\mathbf{\eta }(\xi ,t)+\mathbf{%
\zeta }(\xi ,t).  \label{NPP eq}
\end{equation}%
Here $\mathbf{s}{(\xi ,t)}$ is the position vector of the line points
labelled by the arc length ${\xi }$, which varies in the interval $0<\xi
<l,\ \ $. The quantity $\mathbf{\eta }(\xi ,t)$ stands for dissipation,
which acts for marginally small scales; and the quantity $\mathbf{s}^{\prime
}{(\xi ,t)}$ is the tangent vector. The stirring Langevin force\textbf{\ }$%
\mathbf{\zeta }(\xi ,t)$ is assumed to be Gaussian with a space correlator
concentrated at large scales (of order of the loop size $l$). Explicit forms
of both\textsl{\ }$\eta (\xi ,t)$\textsl{\ }and\textsl{\ }$\zeta (\xi ,t)$%
\textsl{\ }are not essential, since their action is realized at the extreme
ends of the interval, namely at very small and large scales, respectively.
In the intermediate region, the so called the inertial interval, the
influence of both $\eta (\xi ,t)$\textsl{\ }and\textsl{\ }$\zeta (\xi ,t)$%
\textsl{\ }is imperceptible and everything is determined by the Biot-Savart
law, as expressed by the first term on the right-hand side of Eq. (\ref{NPP
eq}). This term controls nonlinear processes and determines spectral
characteristics and flux of energy The role of the Langevin force $\zeta
(\xi ,t)$\ and dissipation $\eta (\xi ,t)$\ \ is reduced to the creation of
additional length (energy) in the large scale region and dissipation of this
excessive length in very small scales. That implies that the length
generated by random force is transferred along the spectrum to be dissipated
at small scales.

\begin{figure}[tbp]
\includegraphics[width=8 cm]{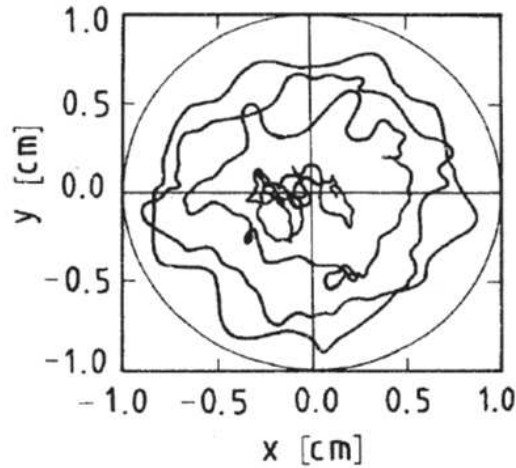}
\caption{ Evolution of a vortex ring under the influence of an external
random force in the local approach (Eq. \protect\ref{NPP eq}); as predicted,
a consequent emergence of higher harmonics takes place leading eventually to
an entanglement of the initially smooth vortex loop. }
\label{loopdeformation}
\end{figure}

An example of the evolution of an initially smooth vortex ring, which obeys
the equation (\ref{NPP eq}), obtained in direct numerical simulation (see
\cite{Nemirovskii1991}), is depicted in Fig. \ref{loopdeformation}.

An analytical study was performed in \cite{Nemirovskii2001b} \ on the basis
of the so-called local induction approximation (LIA), which allows us to
reformulate the Biot-Savart integral in Eq. (\ref{NPP eq}) in a local
nonlinear form. The reduced LIA\ problem in Eq. (\ref{NPP eq}) can be
formulated in the $p$ space, where $p$ is the one-dimensional wave vector,
conjugated to the variable $\xi $. In this form it has an analytical
solution \cite{Nemirovskii2001b}, based on a theoretical trick known as the
direct interaction approximation (DIA) for diagramming technique, elaborated
for classical turbulence by Wyld \cite{Wyld1961}.

One of the results arising from this solution concerns the energy
conservation in the $p$ space, which obviously coincides with the space of
scales (and which to some extent is equivalent the three-dimensional space
of scales)

\begin{equation}
{\frac{{\partial E_{p}}}{{\partial t}}}~+~{\frac{{\partial }P_{p}^{E}}{{%
\partial p}}}~=I_{+}(p)~-~I_{-}(p).  \label{e:a1}
\end{equation}%
Here $E_{p}={\frac{1}{\sqrt{2\pi }}}\int_{0}^{l}\mathbf{s}_{p}^{\prime }%
\mathbf{s}_{-p}^{\prime }e^{-ip\xi }d\xi $ is the one dimensional energy
spectrum in the LIA (up to factor $\rho _{s}\kappa ^{2}$), and $P_{p}^{E}$
is the flux of energy in Fourier space. The right-hand side of Eq. (\ref%
{e:a1}) describes the creation of additional length (energy) at a rate of $%
I_{+}^{K}(p)$ due to the stirring force and its annihilation due to the
dissipative mechanism (at a rate of $-I_{-}^{K}(p)$). In a region of wave
numbers $p$, remote from both the scale of pumping $p_{+}$ and the sink $%
p_{-},\;\;p_{+}~\ll ~p~\ll ~p_{-}$ , the so called inertial interval, the
derivative (in stationary situation) $\partial P_{p}^{E}/\partial p=0$ , so $%
P_{p}^{E}$ is constant. It has therefore been shown that the evolution of a
quantized vortex filament due to its \textquotedblleft
own\textquotedblright\ nonlinear dynamics results in a constant flux of
energy in the space of harmonics \emph{\ }$p$\emph{.}

As in the previous subsection we assume that small $p$ harmonics,
corresponding to large scales (in space of the label variable $\xi $) induce
a velocity field associated with large scale motion in real
three-dimensional space. Accordingly, large\textsl{\ }$p$ harmonics generate
a small scale three-dimensional motion. Therefore the nonlinear exchange of
energy between different harmonics can be associated with the
Kolmogorv-Richardson flux of energy in a real turbulent flow.

Based on the results presented in this section, it can be argued that the
chaotic dynamics of vortex filaments indeed results in a constant flux of
energy of three-dimensional flow, which is the main feature of turbulence.
Thus, in the question \textquotedblleft dog or tail?\textquotedblright ,
these examples weigh in favor of the \textquotedblleft dog\textquotedblright
.

\section{Kolmogorov-type spectra generated by vortex filaments}

Probably, the strongest argument in favour of the quasi-classical behavior
of QT is the $k$ dependence ($k=\left\vert \mathbf{k}\right\vert $, $\mathbf{%
k}$ is the wave number) of the energy spectra$\ E(k)$ of the
three-dimensional velocity field, induced by quantized vortices. Many
numerical simulations, which have been performed both for vortex filament
method for superfluid helium (see, e.g. \cite{Araki2002},\cite%
{Kivotides2002,Kivotides2001c}, \cite{Procaccia2008}) and with the use of
the Gross-Pitaevskii equation for Bose-Einstein condensate (\cite%
{Nore1997,Nore1997a}, \cite{Kobayashi2005},\cite{Tsubota2009a,Sasa2011},
\cite{Nemirovskii2014b}), have demonstrated a Kolmogorov-type dependence $%
E(k)\varpropto \,k^{-5/3}$. The origin of this phenomenon in terms of vortex
line dynamics\ is unclear, details are hidden behind a numerical procedure.

We now discuss how the dynamics of vortex filaments can result in a
Kolmogorov type spectrum. The energy $E$\ of the vortex line configuration $%
\{\mathbf{s}(\xi )\}$ is an integral over the wave vectors $\mathbf{k}$ (see
\cite{Nemirovskii2002}),%
\begin{equation}
E(\mathbf{k)}=\frac{{\rho }_{s}{\kappa }^{2}}{2(2\pi )^{3}}\int\limits_{%
\mathbf{k}}\frac{d^{3}\mathbf{k}}{\mathbf{k}^{2}}\int\limits_{0}^{L}\int%
\limits_{0}^{L}\mathbf{s}^{\prime }(\xi _{1})\cdot \mathbf{s}^{\prime }(\xi
_{2})d\xi _{1}d\xi _{2}\exp \left[ i\int\limits_{\xi _{1}}^{\xi _{2}}\mathbf{%
k\cdot s}^{\prime }(\tilde{\xi})d\tilde{\xi}\right] .  \label{E(k) single}
\end{equation}%
Note that the energy spectrum $E\left( \mathbf{k}\right) $\ can be easily
derived from the expression for the Fourier transform of the velocity field $%
v_{\mathbf{k}}=$\ $\mathbf{k}\times \mathbf{\omega }_{\mathbf{k}}/k^{2}$.
Taking into account that $\mathbf{k\cdot \omega }_{\mathbf{k}}=0$, and that
the vorticity field $\mathbf{\omega (r)}$\ can be written as
\begin{equation}
\mathbf{\omega }(\mathbf{r})=\nabla \times \mathbf{v}_{s}=\kappa \int
\mathbf{s}^{\prime }(\xi )\;\delta (\mathbf{r}-\mathbf{s}(\xi ,t))d\xi ,
\label{vorticity field}
\end{equation}%
one easily obtains equation (\ref{E(k) single}){\large . }In the isotropic
case, the spectral density depends on the absolute value of the wave number $%
k$. Integrating over solid angle leads to formula (see \cite{Kondaurova2005}%
):
\begin{equation}
E(k)=\frac{\rho _{s}\kappa ^{2}}{(2\pi )^{2}}\int\limits_{0}^{L}\int%
\limits_{0}^{L}\mathbf{s}^{\prime }(\xi _{1})\cdot \mathbf{s}^{\prime }(\xi
_{2})\frac{\sin (k\left\vert \mathbf{s}(\xi _{1})-\mathbf{s}(\xi
_{2})\right\vert )}{k\left\vert \mathbf{s}(\xi _{1})-\mathbf{s}(\xi
_{2})\right\vert }d\xi _{1}d\xi _{2}.  \label{E(k) spherical single}
\end{equation}%
{\large \ }

Equations (\ref{E(k) single})-(\ref{E(k) single}) are key relations, which
relate the vortex line configuration $\{\mathbf{s}(\xi )\}$ with the
spectrum of energy. On this ground there was investigated a number of
various configurations of a set of vortex filament appeared on superfluid
flows (see brief review \cite{Nemirovskii2013a}). These results were
obtained using a straight line and vortex ring. We also studied uniform and
non-uniform vortex arrays, a fractal vortex filament, a straight line with
excited Kelvin waves on it, and the case of reconnecting vortex filaments.%
\textsl{\ }Among of the listed above cases there is two configurations $\{%
\mathbf{s}(\xi )\}$,which indeed lead to the Kolmogorov spectrum. These are
the reconnecting vortex filaments \ and Kelvin waves on straight (smooth)
line. Let's follow how they generate the spectrum $E(k)\varpropto \,k^{-5/3}$
and give some comments

\subsection{Reconnecting Vortex Filaments}

During evolution of the quantum turbulence, vortex filaments undergo an
enormous number of collisions and reconnection events. The dynamics of the
line in the process of reconnection possesses many features that have a
universal character (see e.g.\textsl{\ }\cite{Waele1994}).\textsl{\ }This
occurs as a collapse of the approaching vortex filaments, concentrating
large amounts of energy at the points of connection of the line. The idea
that collapsing singular solutions can play a significant role in formation
of turbulent spectra is currently being intensively discussed (see e.g. \cite%
{Kuznetsov2000}, \cite{Kerr2013}). The classical examples of this type of
spectra are the Phillips spectrum for water-wind waves, created by white
caps/wedges on the surface of water, and the Kadomtsev-Petviashvili spectrum
for acoustic turbulence, created by shock waves.\textsl{\ }An attractive
direction for research is to elaborate this idea for the case of
reconnection of vortex filaments in superfluids. In our view, the situation
discussed below is a further example confirming the value of this idea.

\begin{figure}[tbp]
\includegraphics[width=11 cm]{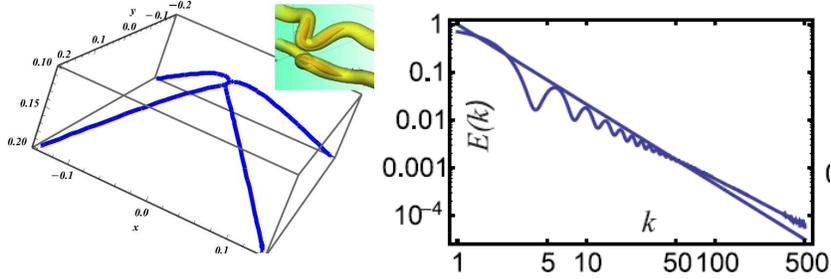}
\caption{(Color online) Left. Touching quasi-hyperbolae describing
collapsing lines \protect\cite{Boue2013}. The inset shows, as an example,
the kinks on the anti-parallel collapsing vortex tubes obtained from
numerical simulation \protect\cite{Bustamante2008}. The figure from numerics
of Bustamante and Kerr is taken to illustrate that the both classical and
quantum vortices implement reconnection in very similar manner%
{\protect\large . }Right. The energy spectrum $E(k)$ of the
three-dimensional velocity field, induced by this configuration. }
\label{reconspertum}
\end{figure}

Over a series of publications (see e.g. \cite{Siggia1985},\cite{Waele1994},
\cite{Kerr2013},\cite{Boue2013},\cite{Andryushchenko2017}), it has been
shown that very near to the point of collision, the vortex filaments have a
universal form (see. Fig. \ref{reconspertum}, left). An analytical
expression for the shape of the curves was proposed in  \cite{Boue2013}.
This expression for the two lines $\{s(\xi _{1})\},\{s(\xi _{2})\}$\ was
used by the author \cite{Nemirovskii2014b} to calculate an energy spectrum $%
E(k)$ on the basis of (angle averaged) Eq. (\ref{E(k) single}). An
investigation was carried out using a method of asymptotic expansion of
integrals containing rapidly oscillating functions, which arise in equations
(\ref{E(k) single})-(\ref{E(k) single}) (see e.g. \cite{Fedoryuk1977}).

From Fig. \ref{reconspertum} (Right) it can be seen that for reconnecting
lines the spectrum is close to $E(k)\propto k^{-5/3}$. The interval of wave
numbers in which the spectrum $E(k)$ $\approx k^{-5/3}$ (straight line) is
observed is regulated by the curvature of the kink and the intervortex space
$\delta $, which is chosen to be equal to unity. In reality this spectrum
covers a maximum of about $\ 1.5$ decades around $k\approx $ $2\pi /\delta $%
. It should be stressed, however, that in the key numerical works (see the
references above),\ the ranges for wave number are also of the order of one
decade around $k\approx $ $2\pi /\delta $.

In a real vortex tangle, where the number of reconnection is large, the
quantity $\delta $\ can vary for different events, This can lead to an
effective enlarging of the interval over which the Kolmogorov spectrum is
realized.

\subsection{1D Kelvin waves spectrum and 3D velocity spectrum}

In the literature there is discussed the idea of obtaining the 3D velocity
spectrum just by putting it equal to the energy spectrum $E_{KW}(k)$ of 1D
Kelvin waves. For instance, as stated in \cite{Kivotides2001b}:"We notice
that, because the fluctuations of the velocity field are induced by the
Kelvin wave fluctuations on the filaments, it is reasonable \ to expect that
\begin{equation}
E(k)\sim E_{KW}(k).  \label{Ev equal Ehd}
\end{equation}%
The same conjecture was used in papers by L'vov et al.(see e.g., \cite%
{Lvov2008}). Details of this activity can be read in a series of papers by
L'vov, Nazarenko and coauthors \cite{L'vov2007}.\cite%
{Lvov2008,Sasa2011,Lvov2011}

Consider the exact solution to this problem on the basis of the general Eqs.
(\ref{E(k) single})-(\ref{E(k) spherical single}). A straight line (along
the $z$\ axis) with running Kelvin waves on it can be represented in terms
of a vector $s(\xi ,t)$\ in the following form: $s(\xi ,t)=(x(z,t),y(z,t),z)$%
. We introduce the two-dimensional vector $a\rho (z,t)=(x(z,t),y(z,t))$,
where the amplitude $a$\ is assumed to be much smaller than the wave length $%
\lambda $. Substituting it into (\ref{E(k) spherical single}) and expanding
in powers of $a$, we get,%
\begin{eqnarray}
E(k) &=&\frac{{\rho }_{s}{\kappa }^{2}}{4\pi ^{2}}\int\limits_{0}^{L}\int%
\limits_{0}^{L}dz_{1}dz_{2}{\LARGE \{}\frac{\sin \left( k\left\vert
z_{2}-z_{1}\right\vert \right) }{k\left\vert z_{2}-z_{1}\right\vert }+\frac{%
a^{2}\cos \left( k\left\vert z_{2}-z_{1}\right\vert \right) (\mathbf{\rho }%
(z_{2})-\mathbf{\rho }(z_{1}))^{2}}{2\left\vert z_{2}-z_{1}\right\vert ^{2}}
\notag \\
&&-\frac{a^{2}\sin \left( k\left\vert z_{2}-z_{1}\right\vert \right) (%
\mathbf{\rho }(z_{2})-\mathbf{\rho }(z_{1}))^{2}}{2k\left\vert
z_{2}-z_{1}\right\vert ^{3}}+\frac{a^{2}(\mathbf{\rho }^{\prime
}(z_{1})\cdot \mathbf{\rho }^{\prime }(z_{2})\sin \left( k\left\vert
z_{2}-z_{1}\right\vert \right) }{k\left\vert z_{2}-z_{1}\right\vert }{\LARGE %
\}}
\end{eqnarray}%
The first term of the zero-order in amplitude $a$\ exactly coincides with
the energy spectrum induced by the unperturbed straight line, as it should
be. Correspondingly it gives $E_{straight}(k)\sim \rho _{s}\kappa ^{2}/k$.

To move further we have to find the correlation characteristics for the
fluctuating vector of displacement $\rho (z_{2})$. The main contribution in
the theory of stochastic nonlinear Kelvin waves and their role in superfluid
turbulence was made in a series of works by Svistunov \cite{Svistunov1995},
Kozik \& Svistunov \cite{Kozik2004,Kozik2005,Kozik2009,Kozik2010} and in
papers by L'vov \& Nazarenko with coauthors \cite%
{Lvov2008,Laurie2010,Lvov2011,Lebedev2010}. Following these works we accept
that the Kelvin waves ensemble has the following power-like spectrum:

\begin{equation}
\left\langle \mathbf{\rho }(p)\cdot \mathbf{\rho }(-p)\right\rangle =Ap^{-s}.
\label{KW- E(k) spectrum}
\end{equation}%
We take here the notation $p$\ for the one-dimensional vector, conjugated to
$z$, reserving the notation $k$\ for the absolute value of the wave vector
of the 3D field. The formula (\ref{KW- E(k) spectrum}) implies that (see,
e.g., \cite{Frisch1995}, Eqs. (4.60),(4.61)) the squared increment for the
vector of displacement scales as, $\left\langle (\mathbf{\rho }(z_{2})-%
\mathbf{\rho }(z_{1}))^{2}\right\rangle \varpropto (z_{2}-z_{1})^{s-1}$.
Then the second order correlator $\left\langle (\mathbf{\rho }^{\prime
}(z_{2})\mathbf{\rho }^{\prime }(z_{1}))\right\rangle $\ scales as $%
\left\langle (\mathbf{\rho }^{\prime }(z_{2})\mathbf{\rho }^{\prime
}(z_{1}))\right\rangle \varpropto (z_{2}-z_{1})^{s-3}$. Substituting it into
(\ref{KW- E(k) spectrum}) and counting the powers of quantity $k$, we conclude
that the correction $E_{KW}(k)$\ to the spectrum $E(k)$,\ due to the
ensemble of Kelvin waves has a form:%
\begin{equation}
E_{KW}(k)\varpropto {a}^{2}k^{-s+2}.  \label{dE due to KW}
\end{equation}%
It is remarkable fact that this quantity coincides formally with the
one-dimensional spectrum of KW $\delta E(p)\varpropto a^{2}p^{-s+2}$. In
series papers L'vov, Nazarenko and coauthors \cite%
{Lvov2008,Sasa2011,L'vov2007,Lvov2011}it was proposed the spectrum for
Kelvin waves of shape (\ref{KW- E(k) spectrum}) with $s=11/3$, therefore the
3D spectrum of velocity field $\delta E(k)\varpropto k^{-5/3}$, i.e. the
Kolmogorov type energy spectrum.

Thus, the 3D motion of superfluid helium induced by chaotic nonlinear Kelvin
waves, indeed possesses the Kolmogorov type spectrum. The contribution $%
E_{KW}(k)$\ is small in comparison with the energy accumulated in the term $%
E_{straight}(k)$, by virtue of the smallness of the wave amplitudes $a$, and
disappears with the KW. There is a view, however, that just like in the 2D
case, when the "own" energy of point vortices, no matter how huge, does not
affect the collective dynamics of the vortex ensemble \cite{Nazarenko2013}.

In this section, it was demonstrated that the dynamics of the lines generate
the vortex

filament configurations $\{\mathbf{s}(\xi )\}$, which induce a
three-dimensional velocity field with a

Kolmogorov-type spectrum. Thus, again, in the question of whether the
vortices are the \textquotedblleft dog\textquotedblright\ or the
\textquotedblleft tail\textquotedblright , these examples weigh in favor of
the \textquotedblleft dog\textquotedblright .

\textsl{\ }

\section{Free decay of quantum turbulence.}

In this section, we consider another line of evidence regarding the
quasi-classical character of quantum turbulence, which relates to the free
decay of QT. This subject is of particular interest, since it has been
properly studied using experimental measurements. Arguments have been put
forward that if quantum turbulence behaves in a similar way to classic
turbulence, then the long-term dependence of the VLD $\mathcal{L}(t)$ is the
power-like function $t^{-3/2}$ \cite{Skrbek2012},\cite{Vinen2010},\cite%
{Bradley2006},\cite{Walmsley2007}. The basis for this result was the belief
that at large scales, the flow of the superfluid component obeys the laws of
classical turbulence. This results in the obsevation that the dissipation
rate $\varepsilon $\ (coinciding with flux of energy $P^{E}$, see Eq. (\ref%
{diss rate}) ) scales as $t^{-3}$, $\varepsilon \propto t^{-3}$. Combining
this result with Eq. (\ref{diss rate}), we arrive at $L(t)\propto t^{-3/2}$.

\begin{figure}[tbp]
\includegraphics[width=14cm]{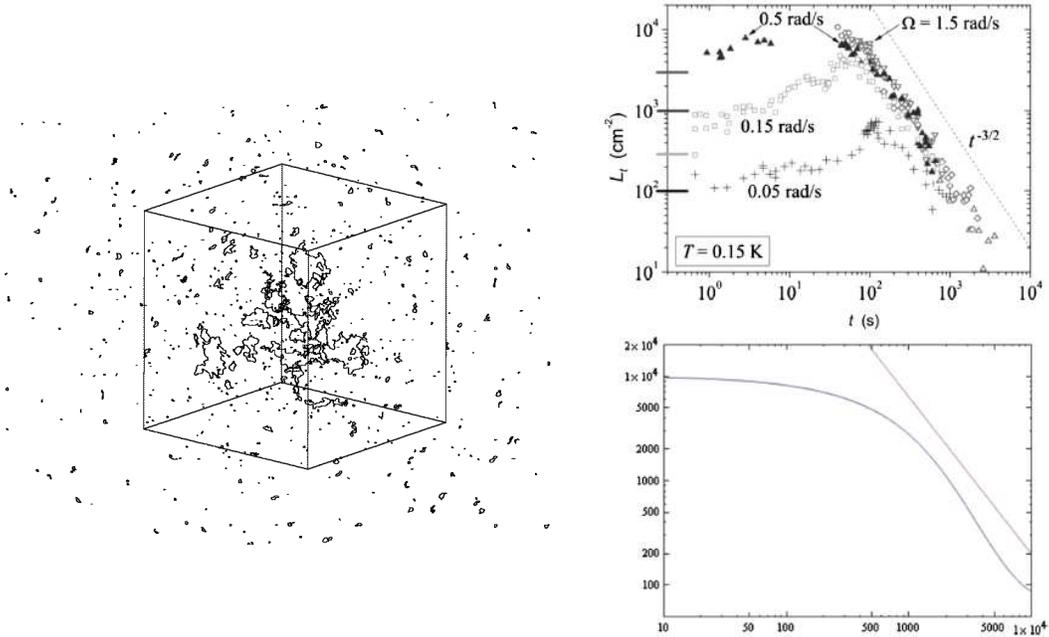}
\caption{Left: The emission of small loops from the bulk; results of direct
numerical simulation (see, \protect\cite{Kondaurova2012}). Right: Comparison
with experiment; the upper section shows the temporal attenuation of the
vortex line density, obtained from experiment \protect\cite{Walmsley2007}.
The temporal behavior of the same quantity calculated on the basis of the
diffusion mechanism in (Eq.\protect\ref{Dif L}) \protect\cite%
{Nemirovskii2010}.}
\label{evapcompar}
\end{figure}

There is a number of experimental works in which this dependence is indeed
observed. One convincing result of an experiment of this type, performed by
the Manchester group \cite{Walmsley2007} is depicted in the upper right-hand
section of Fig \ref{evapcompar}. It can clearly be seen that the long-term
dependence is indeed $\mathcal{L}(t)\propto t^{-3/2}$.

Again, despite the exposed above ideas on the origin of the dependence $%
\mathcal{L}(t)\propto t^{-3/2}$, it is not clear how this arises directly
from the dynamics of vortex lines.

We now discuss the mechanism of free decay of quantum turbulence which is
based on the dynamics of vortex filaments and which predicts the dependence $%
\mathcal{L}(t)\propto t^{-3/2}$.\emph{\ }This mechanism is related to the
emission of small loops from the bulk. As discussed earlier, a large number
of very small loops appear during the reconnection cascade. However, due to
their small size, these loops have high mobility and escape from the volume.
An example of this behavior, obtained by direct numerical simulation (see,
\cite{Kondaurova2012}), is shown in the left-hand section of Fig. \ref%
{evapcompar}. This process occurs in a diffusion-like manner.

The quantitative theory of the diffusion-like spreading of the VT, with the
consequent decay of quantum turbulence, was developed by the author, and
details can be found in \cite{Nemirovskii2010}. The vortex loops composing
the vortex tangle can move as a whole, with a drift velocity $V_{l}$
depending on their structure and their length. The loops moving in space
carry the length of the line, the energy, the momentum, and so on.\emph{\ }%
In the case of an inhomogeneous vortex tangle, the net flux $\mathbf{J}$ of
the vortex length appears due to the gradient of concentration of the vortex
line density $\mathcal{L}(x,t)$. The situation here is exactly the same as
in classical kinetic theory, with the difference being that the
\textquotedblleft carriers\textquotedblright\ are not point particles but
extended objects (vortex loops), which themselves possess an infinite number
of degrees of freedom with very involved dynamics. It is possible to carry
out an investigation based on the supposition that vortex loops have a
Brownian or random walk structure with a generalized Wiener distribution. To
develop a theory of transport processes fulfilled by vortex loops (in the
spirit of classical kinetic theory), we need to calculate the drift velocity
$V_{l}$ and the free path $\lambda (l)$\ for a loop of size $l$. The free
path $\lambda (l)$ is very small, implying that only very small loops make a
significant contribution to transport processes, e.g. a loop of size equal
to the interline space $\delta =\mathcal{L}^{-1/2}$ can fly (on average)
about $2-3$ times the distance $\delta $. However, when the vortex tangle
becomes sufficiently dilute, the larger loops can move a distance comparable
with the size of the bulk. Referring to \cite{Nemirovskii2010} we write down
here the final result. The flux $\mathbf{J}$ of vortex lines is proportional
to $\nabla \mathcal{L}$; $\mathbf{J}=D_{v}\nabla \mathcal{L}$; and,
correspondingly, the spatial-temporal evolution of quantity $\mathcal{L}$
obeys the diffusion-type equation%
\begin{equation}
\frac{\partial \mathcal{L}}{\partial t}=D_{v}\nabla ^{2}\mathcal{L}.
\label{Dif L}
\end{equation}

Here the diffusion constant $D_{v}\approx 2\kappa $. To clarify whether this
mechanism is able to sustain the decay of the QT observed in the Manchester
experiments \cite{Walmsley2007}, we calculated the evolution of the VLD $%
\mathcal{L}(t)$\ solving the diffusion equation, for the cases corresponding
to these experimental conditions. The results are shown in the lower
right-hand section of Fig. \ref{evapcompar}.\ It is evident that this
theoretical result based on the diffusion model, is in very good agreement
with the experimental data.

Thus, based on a theoretical analysis of the vortex line dynamics, we have
demonstrated how processes occurring in the ensemble of vortex loops lead to
the behavior observed in classical turbulence that again supports the
\textquotedblleft dog\textquotedblright\ concept.

\section{Conclusion and Discussion}

We therefore demonstrate that certain results obtained for quantum fluids,
which have been considered to be arguments in favor of the idea of modeling
classical turbulence using a set of chaotic quantized vortices, can be
explained using the frame of theoretical models dealing with the dynamics of
vortex filaments.

Of course, one can object that the current arguments are not strong enough
to claim that the issue has been definitively settled, and that the
phenomena described above have nothing to do with the real turbulence.
Nevertheless, taking into account the tantalizing problem of classical
turbulence, the hope of solving it using a new breakthrough approach seems
very attractive. Thus, the following bifurcation arise. An optimistic view
is that these approaches do indeed relate to the real turbulence, and that
we have new insight into processes occurring in turbulent flow. A
pessimistic view is that the phenomena described above have nothing to do
with the real turbulence, and that this effect is a coincidence. However, we
hope in any event that this paper will have a stimulating effect on related
activity.

The study of Sections  II-A, III, IV was carried out under state contract
with IT SB RAS (АААА-А17-117022850027-5), the study of subsections II-B III-B was
financially supported by RFBR / Russian Science Foundation (Project No.
18-08-00576).


\begin{thebibliography}{56}
\expandafter\ifx\csname natexlab\endcsname\relax\def\natexlab#1{#1}\fi
\expandafter\ifx\csname bibnamefont\endcsname\relax
  \def\bibnamefont#1{#1}\fi
\expandafter\ifx\csname bibfnamefont\endcsname\relax
  \def\bibfnamefont#1{#1}\fi
\expandafter\ifx\csname citenamefont\endcsname\relax
  \def\citenamefont#1{#1}\fi
\expandafter\ifx\csname url\endcsname\relax
  \def\url#1{\texttt{#1}}\fi
\expandafter\ifx\csname urlprefix\endcsname\relax\def\urlprefix{URL }\fi
\providecommand{\bibinfo}[2]{#2}
\providecommand{\eprint}[2][]{\url{#2}}

\bibitem[{\citenamefont{Frisch}(1995)}]{Frisch1995}
\bibinfo{author}{\bibfnamefont{U.}~\bibnamefont{Frisch}},
  \emph{\bibinfo{title}{Turbulence}} (\bibinfo{publisher}{Cambridge University
  Press, Cambridge}, \bibinfo{year}{1995}).

\bibitem[{\citenamefont{Chorin}(1994)}]{Chorin1994}
\bibinfo{author}{\bibfnamefont{A.}~\bibnamefont{Chorin}},
  \emph{\bibinfo{title}{Vorticity and turbulence}}, Applied mathematical
  sciences (\bibinfo{publisher}{Springer-Verlag}, \bibinfo{year}{1994}).

\bibitem[{\citenamefont{Migdal}(1986)}]{Migdal1986}
\bibinfo{author}{\bibfnamefont{A.}~\bibnamefont{Migdal}},
  \bibinfo{journal}{Voprosi Kibernetiki} p. \bibinfo{pages}{122}
  (\bibinfo{year}{1986}).

\bibitem[{\citenamefont{Tsubota and Kobayashi}(2009)}]{Tsubota2009a}
\bibinfo{author}{\bibfnamefont{M.}~\bibnamefont{Tsubota}} \bibnamefont{and}
  \bibinfo{author}{\bibfnamefont{M.}~\bibnamefont{Kobayashi}}, in
  \emph{\bibinfo{booktitle}{Progress in Low TEMPERATURE PHYSICS: QUANTUM
  TURBULENCE}}, edited by
  \bibinfo{editor}{\bibfnamefont{M.}~\bibnamefont{Tsubota}} \bibnamefont{and}
  \bibinfo{editor}{\bibfnamefont{W.}~\bibnamefont{Halperin}}
  (\bibinfo{publisher}{Elsevier}, \bibinfo{year}{2009}),
  vol.~\bibinfo{volume}{16} of \emph{\bibinfo{series}{Progress in Low
  Temperature Physics}}, pp. \bibinfo{pages}{1 -- 43}.

\bibitem[{\citenamefont{Vincent and Meneguzzi}(1991)}]{Vincent1991}
\bibinfo{author}{\bibfnamefont{A.}~\bibnamefont{Vincent}} \bibnamefont{and}
  \bibinfo{author}{\bibfnamefont{M.}~\bibnamefont{Meneguzzi}},
  \bibinfo{journal}{Journal of Fluid Mechanics} \textbf{\bibinfo{volume}{225}},
  \bibinfo{pages}{1} (\bibinfo{year}{1991}).

\bibitem[{\citenamefont{Vinen}(2000)}]{Vinen2000}
\bibinfo{author}{\bibfnamefont{W.~F.} \bibnamefont{Vinen}},
  \bibinfo{journal}{Phys. Rev. B} \textbf{\bibinfo{volume}{61}},
  \bibinfo{pages}{1410} (\bibinfo{year}{2000}).

\bibitem[{\citenamefont{Kobayashi and Tsubota}(2005)}]{Kobayashi2005}
\bibinfo{author}{\bibfnamefont{M.}~\bibnamefont{Kobayashi}} \bibnamefont{and}
  \bibinfo{author}{\bibfnamefont{M.}~\bibnamefont{Tsubota}},
  \bibinfo{journal}{Phys. Rev. Lett.} \textbf{\bibinfo{volume}{94}},
  \bibinfo{pages}{065302} (\bibinfo{year}{2005}).

\bibitem[{\citenamefont{Vinen}(2010)}]{Vinen2010}
\bibinfo{author}{\bibfnamefont{W.}~\bibnamefont{Vinen}},
  \bibinfo{journal}{Journal of Low Temperature Physics}
  \textbf{\bibinfo{volume}{161}}, \bibinfo{pages}{419} (\bibinfo{year}{2010}).

\bibitem[{\citenamefont{Skrbek and Sreenivasan}(2012)}]{Skrbek2012}
\bibinfo{author}{\bibfnamefont{L.}~\bibnamefont{Skrbek}} \bibnamefont{and}
  \bibinfo{author}{\bibfnamefont{K.~R.} \bibnamefont{Sreenivasan}},
  \bibinfo{journal}{Physics of Fluids} \textbf{\bibinfo{volume}{24}},
  \bibinfo{eid}{011301} (pages~\bibinfo{numpages}{48}) (\bibinfo{year}{2012}).

\bibitem[{\citenamefont{Tsubota et~al.}(2013)\citenamefont{Tsubota, Kobayashi,
  and Takeuchi}}]{Tsubota2013}
\bibinfo{author}{\bibfnamefont{M.}~\bibnamefont{Tsubota}},
  \bibinfo{author}{\bibfnamefont{M.}~\bibnamefont{Kobayashi}},
  \bibnamefont{and} \bibinfo{author}{\bibfnamefont{H.}~\bibnamefont{Takeuchi}},
  \bibinfo{journal}{Physics Reports} \textbf{\bibinfo{volume}{522}},
  \bibinfo{pages}{191 } (\bibinfo{year}{2013}), ISSN \bibinfo{issn}{0370-1573},
  \bibinfo{note}{quantum hydrodynamics}.

\bibitem[{\citenamefont{Barenghi et~al.}(2014)\citenamefont{Barenghi, Skrbek,
  and Sreenivasan}}]{Barenghi2014a}
\bibinfo{author}{\bibfnamefont{C.~F.} \bibnamefont{Barenghi}},
  \bibinfo{author}{\bibfnamefont{L.}~\bibnamefont{Skrbek}}, \bibnamefont{and}
  \bibinfo{author}{\bibfnamefont{K.~R.} \bibnamefont{Sreenivasan}},
  \bibinfo{journal}{Proceedings of the National Academy of Sciences}
  \textbf{\bibinfo{volume}{111}}, \bibinfo{pages}{4647} (\bibinfo{year}{2014}).
 

\bibitem[{\citenamefont{Walmsley et~al.}(2007)\citenamefont{Walmsley, Golov,
  Hall, Levchenko, and Vinen}}]{Walmsley2007}
\bibinfo{author}{\bibfnamefont{P.~M.} \bibnamefont{Walmsley}},
  \bibinfo{author}{\bibfnamefont{A.~I.} \bibnamefont{Golov}},
  \bibinfo{author}{\bibfnamefont{H.~E.} \bibnamefont{Hall}},
  \bibinfo{author}{\bibfnamefont{A.~A.} \bibnamefont{Levchenko}},
  \bibnamefont{and} \bibinfo{author}{\bibfnamefont{W.~F.} \bibnamefont{Vinen}},
  \bibinfo{journal}{Phys. Rev. Lett.} \textbf{\bibinfo{volume}{99}},
  \bibinfo{pages}{265302} (\bibinfo{year}{2007}).

\bibitem[{\citenamefont{Kleinert}(1990)}]{Kleinert1990}
\bibinfo{author}{\bibfnamefont{H.}~\bibnamefont{Kleinert}},
  \emph{\bibinfo{title}{Gauge Fields in Condenced Matter Physics}}
  (\bibinfo{publisher}{World Scientific, Singapore}, \bibinfo{year}{1990}).

\bibitem[{\citenamefont{Nemirovskii}(1998)}]{Nemirovskii1998}
\bibinfo{author}{\bibfnamefont{S.~K.} \bibnamefont{Nemirovskii}},
  \bibinfo{journal}{Phys. Rev. B} \textbf{\bibinfo{volume}{57}},
  \bibinfo{pages}{5972} (\bibinfo{year}{1998}).

\bibitem[{\citenamefont{Nemirovskii}(2006)}]{Nemirovskii2006}
\bibinfo{author}{\bibfnamefont{S.~K.} \bibnamefont{Nemirovskii}},
  \bibinfo{journal}{Phys. Rev. Lett.} \textbf{\bibinfo{volume}{96}},
  \bibinfo{pages}{015301} (\bibinfo{year}{2006}).

\bibitem[{\citenamefont{Nemirovskii}(2008)}]{Nemirovskii2008}
\bibinfo{author}{\bibfnamefont{S.~K.} \bibnamefont{Nemirovskii}},
  \bibinfo{journal}{Phys. Rev. B} \textbf{\bibinfo{volume}{77}},
  \bibinfo{pages}{214509} (\bibinfo{year}{2008}).

\bibitem[{\citenamefont{Nemirovskii}(2013{\natexlab{a}})}]{Nemirovskii2013}
\bibinfo{author}{\bibfnamefont{S.~K.} \bibnamefont{Nemirovskii}},
  \bibinfo{journal}{Physics Reports} \textbf{\bibinfo{volume}{524}},
  \bibinfo{pages}{85 } (\bibinfo{year}{2013}{\natexlab{a}}).

\bibitem[{\citenamefont{Zakharov et~al.}(1992)\citenamefont{Zakharov, L’vov,
  and Falkovich}}]{Zakharov1992}
\bibinfo{author}{\bibfnamefont{V.~E.} \bibnamefont{Zakharov}},
  \bibinfo{author}{\bibfnamefont{V.~S.} \bibnamefont{L’vov}}, \bibnamefont{and}
  \bibinfo{author}{\bibfnamefont{G.}~\bibnamefont{Falkovich}},
  \emph{\bibinfo{title}{Kolmogorov Spectra of Turbulence I}}
  (\bibinfo{publisher}{Springer-Verlag, Berlin}, \bibinfo{year}{1992}).

\bibitem[{\citenamefont{Feynman}(1955)}]{Feynman1955}
\bibinfo{author}{\bibfnamefont{R.~P.} \bibnamefont{Feynman}},
  \emph{\bibinfo{title}{Progress in Low Temperature Physics, Vol. 1}}
  (\bibinfo{publisher}{North-Holland, Amsterdam}, \bibinfo{year}{1955}),
  p.~\bibinfo{pages}{17}.

\bibitem[{\citenamefont{Nemirovskii and Baltsevich}(2001)}]{Nemirovskii2001b}
\bibinfo{author}{\bibfnamefont{S.}~\bibnamefont{Nemirovskii}} \bibnamefont{and}
  \bibinfo{author}{\bibfnamefont{A.}~\bibnamefont{Baltsevich}}, in
  \emph{\bibinfo{booktitle}{Quantized Vortex Dynamics and Superfluid
  Turbulence}}, edited by
  \bibinfo{editor}{\bibfnamefont{C.}~\bibnamefont{Barenghi}},
  \bibinfo{editor}{\bibfnamefont{R.}~\bibnamefont{Donnelly}}, \bibnamefont{and}
  \bibinfo{editor}{\bibfnamefont{W.}~\bibnamefont{Vinen}}
  (\bibinfo{publisher}{Springer Berlin Heidelberg}, \bibinfo{year}{2001}), vol.
  \bibinfo{volume}{571} of \emph{\bibinfo{series}{Lecture Notes in Physics}},
  pp. \bibinfo{pages}{219--225}, ISBN \bibinfo{isbn}{978-3-540-42226-6}.

\bibitem[{\citenamefont{Nemirovskii et~al.}(1991)\citenamefont{Nemirovskii,
  Pakleza, and Poppe}}]{Nemirovskii1991}
\bibinfo{author}{\bibfnamefont{S.~K.} \bibnamefont{Nemirovskii}},
  \bibinfo{author}{\bibfnamefont{J.}~\bibnamefont{Pakleza}}, \bibnamefont{and}
  \bibinfo{author}{\bibfnamefont{W.}~\bibnamefont{Poppe}},
  \emph{\bibinfo{title}{Notes et Documents LIMSI}}
  (\bibinfo{publisher}{Laboratoire d'Informatique pour la Mecanique et les
  Sciences de l'Ingenieur) No. 91-14}, \bibinfo{year}{1991}).

\bibitem[{\citenamefont{Wyld}(1961)}]{Wyld1961}
\bibinfo{author}{\bibfnamefont{H.~W.} \bibnamefont{Wyld}},
  \bibinfo{journal}{Annals of Physics} \textbf{\bibinfo{volume}{14}},
  \bibinfo{pages}{143} (\bibinfo{year}{1961}).

\bibitem[{\citenamefont{Araki et~al.}(2002)\citenamefont{Araki, Tsubota, and
  Nemirovskii}}]{Araki2002}
\bibinfo{author}{\bibfnamefont{T.}~\bibnamefont{Araki}},
  \bibinfo{author}{\bibfnamefont{M.}~\bibnamefont{Tsubota}}, \bibnamefont{and}
  \bibinfo{author}{\bibfnamefont{S.~K.} \bibnamefont{Nemirovskii}},
  \bibinfo{journal}{Phys. Rev. Lett.} \textbf{\bibinfo{volume}{89}},
  \bibinfo{pages}{145301} (\bibinfo{year}{2002}).

\bibitem[{\citenamefont{Kivotides et~al.}(2002)\citenamefont{Kivotides,
  Vassilicos, Samuels, and Barenghi}}]{Kivotides2002}
\bibinfo{author}{\bibfnamefont{D.}~\bibnamefont{Kivotides}},
  \bibinfo{author}{\bibfnamefont{C.~J.} \bibnamefont{Vassilicos}},
  \bibinfo{author}{\bibfnamefont{D.~C.} \bibnamefont{Samuels}},
  \bibnamefont{and} \bibinfo{author}{\bibfnamefont{C.~F.}
  \bibnamefont{Barenghi}}, \bibinfo{journal}{EPL (Europhysics Letters)}
  \textbf{\bibinfo{volume}{57}}, \bibinfo{pages}{845} (\bibinfo{year}{2002}).

\bibitem[{\citenamefont{Kivotides
  et~al.}(2001{\natexlab{a}})\citenamefont{Kivotides, Barenghi, and
  Samuels}}]{Kivotides2001c}
\bibinfo{author}{\bibfnamefont{D.}~\bibnamefont{Kivotides}},
  \bibinfo{author}{\bibfnamefont{C.~F.} \bibnamefont{Barenghi}},
  \bibnamefont{and} \bibinfo{author}{\bibfnamefont{D.~C.}
  \bibnamefont{Samuels}}, \bibinfo{journal}{Europhys. Lett.}
  \textbf{\bibinfo{volume}{54}}, \bibinfo{pages}{771}
  (\bibinfo{year}{2001}{\natexlab{a}}).

\bibitem[{\citenamefont{Procaccia and Sreenivasan}(2008)}]{Procaccia2008}
\bibinfo{author}{\bibfnamefont{I.}~\bibnamefont{Procaccia}} \bibnamefont{and}
  \bibinfo{author}{\bibfnamefont{K.}~\bibnamefont{Sreenivasan}},
  \bibinfo{journal}{Physica D: Nonlinear Phenomena}
  \textbf{\bibinfo{volume}{237}}, \bibinfo{pages}{2167 }
  (\bibinfo{year}{2008}), \bibinfo{note}{euler Equations: 250 Years On -
  Proceedings of an international conference}.

\bibitem[{\citenamefont{Nore et~al.}(1997{\natexlab{a}})\citenamefont{Nore,
  Abid, and Brachet}}]{Nore1997}
\bibinfo{author}{\bibfnamefont{C.}~\bibnamefont{Nore}},
  \bibinfo{author}{\bibfnamefont{M.}~\bibnamefont{Abid}}, \bibnamefont{and}
  \bibinfo{author}{\bibfnamefont{M.~E.} \bibnamefont{Brachet}},
  \bibinfo{journal}{Phys. Rev. Lett.} \textbf{\bibinfo{volume}{78}},
  \bibinfo{pages}{3896} (\bibinfo{year}{1997}{\natexlab{a}}).

\bibitem[{\citenamefont{Nore et~al.}(1997{\natexlab{b}})\citenamefont{Nore,
  Abid, and Brachet}}]{Nore1997a}
\bibinfo{author}{\bibfnamefont{C.}~\bibnamefont{Nore}},
  \bibinfo{author}{\bibfnamefont{M.}~\bibnamefont{Abid}}, \bibnamefont{and}
  \bibinfo{author}{\bibfnamefont{M.~E.} \bibnamefont{Brachet}},
  \bibinfo{journal}{Physics of Fluids} \textbf{\bibinfo{volume}{9}},
  \bibinfo{pages}{2644} (\bibinfo{year}{1997}{\natexlab{b}}).

\bibitem[{\citenamefont{Sasa et~al.}(2011)\citenamefont{Sasa, Kano, Machida,
  L'vov, Rudenko, and Tsubota}}]{Sasa2011}
\bibinfo{author}{\bibfnamefont{N.}~\bibnamefont{Sasa}},
  \bibinfo{author}{\bibfnamefont{T.}~\bibnamefont{Kano}},
  \bibinfo{author}{\bibfnamefont{M.}~\bibnamefont{Machida}},
  \bibinfo{author}{\bibfnamefont{V.~S.} \bibnamefont{L'vov}},
  \bibinfo{author}{\bibfnamefont{O.}~\bibnamefont{Rudenko}}, \bibnamefont{and}
  \bibinfo{author}{\bibfnamefont{M.}~\bibnamefont{Tsubota}},
  \bibinfo{journal}{Phys. Rev. B} \textbf{\bibinfo{volume}{84}},
  \bibinfo{pages}{054525} (\bibinfo{year}{2011}).

\bibitem[{\citenamefont{Nemirovskii}(2014)}]{Nemirovskii2014b}
\bibinfo{author}{\bibfnamefont{S.~K.} \bibnamefont{Nemirovskii}},
  \bibinfo{journal}{Physical Review B} \textbf{\bibinfo{volume}{90}},
  \bibinfo{pages}{104506} (\bibinfo{year}{2014}).

\bibitem[{\citenamefont{Nemirovskii et~al.}(2002)\citenamefont{Nemirovskii,
  Tsubota, and Araki}}]{Nemirovskii2002}
\bibinfo{author}{\bibfnamefont{S.~K.} \bibnamefont{Nemirovskii}},
  \bibinfo{author}{\bibfnamefont{M.}~\bibnamefont{Tsubota}}, \bibnamefont{and}
  \bibinfo{author}{\bibfnamefont{T.}~\bibnamefont{Araki}},
  \bibinfo{journal}{Journal of Low Temperature Physics}
  \textbf{\bibinfo{volume}{126}}, \bibinfo{pages}{1535} (\bibinfo{year}{2002}).

\bibitem[{\citenamefont{Kondaurova and Nemirovskii}(2005)}]{Kondaurova2005}
\bibinfo{author}{\bibfnamefont{L.}~\bibnamefont{Kondaurova}} \bibnamefont{and}
  \bibinfo{author}{\bibfnamefont{S.~K.} \bibnamefont{Nemirovskii}},
  \bibinfo{journal}{Journal of Low Temperature Physics}
  \textbf{\bibinfo{volume}{138}}, \bibinfo{pages}{555} (\bibinfo{year}{2005}).

\bibitem[{\citenamefont{Nemirovskii}(2013{\natexlab{b}})}]{Nemirovskii2013a}
\bibinfo{author}{\bibfnamefont{S.}~\bibnamefont{Nemirovskii}},
  \bibinfo{journal}{Journal of Low Temperature Physics}
  \textbf{\bibinfo{volume}{171}}, \bibinfo{pages}{504}
  (\bibinfo{year}{2013}{\natexlab{b}}).

\bibitem[{\citenamefont{de~Waele and Aarts}(1994)}]{Waele1994}
\bibinfo{author}{\bibfnamefont{A.~T. A.~M.} \bibnamefont{de~Waele}}
  \bibnamefont{and} \bibinfo{author}{\bibfnamefont{R.~G. K.~M.}
  \bibnamefont{Aarts}}, \bibinfo{journal}{Phys. Rev. Lett.}
  \textbf{\bibinfo{volume}{72}}, \bibinfo{pages}{482} (\bibinfo{year}{1994}).

\bibitem[{\citenamefont{Kuznetsov and Ruban}(2000)}]{Kuznetsov2000}
\bibinfo{author}{\bibfnamefont{E.}~\bibnamefont{Kuznetsov}} \bibnamefont{and}
  \bibinfo{author}{\bibfnamefont{V.}~\bibnamefont{Ruban}},
  \bibinfo{journal}{Journal of Experimental and Theoretical Physics}
  \textbf{\bibinfo{volume}{91}}, \bibinfo{pages}{775} (\bibinfo{year}{2000}).

\bibitem[{\citenamefont{Kerr}(2013)}]{Kerr2013}
\bibinfo{author}{\bibfnamefont{R.~M.} \bibnamefont{Kerr}},
  \bibinfo{journal}{Physics of Fluids} \textbf{\bibinfo{volume}{25}},
  \bibinfo{eid}{065101} (\bibinfo{year}{2013}).

\bibitem[{\citenamefont{Siggia}(1985)}]{Siggia1985}
\bibinfo{author}{\bibfnamefont{E.~D.} \bibnamefont{Siggia}},
  \bibinfo{journal}{Phys. Fluids} \textbf{\bibinfo{volume}{28}},
  \bibinfo{pages}{794} (\bibinfo{year}{1985}).

\bibitem[{\citenamefont{Bou\'e et~al.}(2013)\citenamefont{Bou\'e, Khomenko,
  L'vov, and Procaccia}}]{Boue2013}
\bibinfo{author}{\bibfnamefont{L.}~\bibnamefont{Bou\'e}},
  \bibinfo{author}{\bibfnamefont{D.}~\bibnamefont{Khomenko}},
  \bibinfo{author}{\bibfnamefont{V.~S.} \bibnamefont{L'vov}}, \bibnamefont{and}
  \bibinfo{author}{\bibfnamefont{I.}~\bibnamefont{Procaccia}},
  \bibinfo{journal}{Phys. Rev. Lett.} \textbf{\bibinfo{volume}{111}},
  \bibinfo{pages}{145302} (\bibinfo{year}{2013}).

\bibitem[{\citenamefont{Andryushchenko
  et~al.}(2017)\citenamefont{Andryushchenko, Kondaurova, and
  Nemirovskii}}]{Andryushchenko2017}
\bibinfo{author}{\bibfnamefont{V.~A.} \bibnamefont{Andryushchenko}},
  \bibinfo{author}{\bibfnamefont{L.~P.} \bibnamefont{Kondaurova}},
  \bibnamefont{and} \bibinfo{author}{\bibfnamefont{S.~K.}
  \bibnamefont{Nemirovskii}}, \bibinfo{journal}{Journal of Low Temperature
  Physics} \textbf{\bibinfo{volume}{187}}, \bibinfo{pages}{523}
  (\bibinfo{year}{2017}), ISSN \bibinfo{issn}{1573-7357}.

\bibitem[{\citenamefont{Fedoryuk}(1977)}]{Fedoryuk1977}
\bibinfo{author}{\bibfnamefont{M.~V.} \bibnamefont{Fedoryuk}},
  \emph{\bibinfo{title}{Method of Saddle Points.}} (\bibinfo{publisher}{Nauka,
  Moscow}, \bibinfo{year}{1977}).

\bibitem[{\citenamefont{Bustamante and Kerr}(2008)}]{Bustamante2008}
\bibinfo{author}{\bibfnamefont{M.~D.} \bibnamefont{Bustamante}}
  \bibnamefont{and} \bibinfo{author}{\bibfnamefont{R.~M.} \bibnamefont{Kerr}},
  \bibinfo{journal}{Physica D: Nonlinear Phenomena}
  \textbf{\bibinfo{volume}{237}}, \bibinfo{pages}{1912 }
  (\bibinfo{year}{2008}).

\bibitem[{\citenamefont{Kivotides
  et~al.}(2001{\natexlab{b}})\citenamefont{Kivotides, Vassilicos, Samuels, and
  Barenghi}}]{Kivotides2001b}
\bibinfo{author}{\bibfnamefont{D.}~\bibnamefont{Kivotides}},
  \bibinfo{author}{\bibfnamefont{J.~C.} \bibnamefont{Vassilicos}},
  \bibinfo{author}{\bibfnamefont{D.~C.} \bibnamefont{Samuels}},
  \bibnamefont{and} \bibinfo{author}{\bibfnamefont{C.~F.}
  \bibnamefont{Barenghi}}, \bibinfo{journal}{Phys. Rev. Lett.}
  \textbf{\bibinfo{volume}{86}}, \bibinfo{pages}{3080}
  (\bibinfo{year}{2001}{\natexlab{b}}).

\bibitem[{\citenamefont{L’vov et~al.}(2008)\citenamefont{L’vov, Nazarenko, and
  Rudenko}}]{Lvov2008}
\bibinfo{author}{\bibfnamefont{V.}~\bibnamefont{L’vov}},
  \bibinfo{author}{\bibfnamefont{S.}~\bibnamefont{Nazarenko}},
  \bibnamefont{and} \bibinfo{author}{\bibfnamefont{O.}~\bibnamefont{Rudenko}},
  \bibinfo{journal}{J. Low Temp. Phys.} \textbf{\bibinfo{volume}{153}},
  \bibinfo{pages}{140} (\bibinfo{year}{2008}).

\bibitem[{\citenamefont{L'vov et~al.}(2007)\citenamefont{L'vov, Nazarenko, and
  Rudenko}}]{L'vov2007}
\bibinfo{author}{\bibfnamefont{V.~S.} \bibnamefont{L'vov}},
  \bibinfo{author}{\bibfnamefont{S.~V.} \bibnamefont{Nazarenko}},
  \bibnamefont{and} \bibinfo{author}{\bibfnamefont{O.}~\bibnamefont{Rudenko}},
  \bibinfo{journal}{Phys. Rev. B} \textbf{\bibinfo{volume}{76}},
  \bibinfo{pages}{024520} (\bibinfo{year}{2007}).

\bibitem[{\citenamefont{Bou\'e et~al.}(2011)\citenamefont{Bou\'e, Dasgupta,
  Laurie, L'vov, Nazarenko, and Procaccia}}]{Lvov2011}
\bibinfo{author}{\bibfnamefont{L.}~\bibnamefont{Bou\'e}},
  \bibinfo{author}{\bibfnamefont{R.}~\bibnamefont{Dasgupta}},
  \bibinfo{author}{\bibfnamefont{J.}~\bibnamefont{Laurie}},
  \bibinfo{author}{\bibfnamefont{V.}~\bibnamefont{L'vov}},
  \bibinfo{author}{\bibfnamefont{S.}~\bibnamefont{Nazarenko}},
  \bibnamefont{and}
  \bibinfo{author}{\bibfnamefont{I.}~\bibnamefont{Procaccia}},
  \bibinfo{journal}{Phys. Rev. B} \textbf{\bibinfo{volume}{84}},
  \bibinfo{pages}{064516} (\bibinfo{year}{2011}).

\bibitem[{\citenamefont{Svistunov}(1995)}]{Svistunov1995}
\bibinfo{author}{\bibfnamefont{B.~V.} \bibnamefont{Svistunov}},
  \bibinfo{journal}{Phys. Rev. B} \textbf{\bibinfo{volume}{52}},
  \bibinfo{pages}{3647} (\bibinfo{year}{1995}).

\bibitem[{\citenamefont{Kozik and Svistunov}(2004)}]{Kozik2004}
\bibinfo{author}{\bibfnamefont{E.}~\bibnamefont{Kozik}} \bibnamefont{and}
  \bibinfo{author}{\bibfnamefont{B.}~\bibnamefont{Svistunov}},
  \bibinfo{journal}{Phys. Rev. Lett.} \textbf{\bibinfo{volume}{92}},
  \bibinfo{pages}{035301} (\bibinfo{year}{2004}).

\bibitem[{\citenamefont{Kozik and Svistunov}(2005)}]{Kozik2005}
\bibinfo{author}{\bibfnamefont{E.}~\bibnamefont{Kozik}} \bibnamefont{and}
  \bibinfo{author}{\bibfnamefont{B.}~\bibnamefont{Svistunov}},
  \bibinfo{journal}{Phys. Rev. Lett.} \textbf{\bibinfo{volume}{94}},
  \bibinfo{pages}{025301} (\bibinfo{year}{2005}).

\bibitem[{\citenamefont{Kozik and Svistunov}(2009)}]{Kozik2009}
\bibinfo{author}{\bibfnamefont{E.}~\bibnamefont{Kozik}} \bibnamefont{and}
  \bibinfo{author}{\bibfnamefont{B.}~\bibnamefont{Svistunov}},
  \bibinfo{journal}{Journal of Low Temperature Physics}
  \textbf{\bibinfo{volume}{156}}, \bibinfo{pages}{215} (\bibinfo{year}{2009}).

\bibitem[{\citenamefont{Kozik and Svistunov}(2010)}]{Kozik2010}
\bibinfo{author}{\bibfnamefont{E.}~\bibnamefont{Kozik}} \bibnamefont{and}
  \bibinfo{author}{\bibfnamefont{B.}~\bibnamefont{Svistunov}},
  \bibinfo{journal}{Phys. Rev. B} \textbf{\bibinfo{volume}{82}},
  \bibinfo{pages}{140510} (\bibinfo{year}{2010}).

\bibitem[{\citenamefont{Laurie et~al.}(2010)\citenamefont{Laurie, L'vov,
  Nazarenko, and Rudenko}}]{Laurie2010}
\bibinfo{author}{\bibfnamefont{J.}~\bibnamefont{Laurie}},
  \bibinfo{author}{\bibfnamefont{V.~S.} \bibnamefont{L'vov}},
  \bibinfo{author}{\bibfnamefont{S.}~\bibnamefont{Nazarenko}},
  \bibnamefont{and} \bibinfo{author}{\bibfnamefont{O.}~\bibnamefont{Rudenko}},
  \bibinfo{journal}{Phys. Rev. B} \textbf{\bibinfo{volume}{81}},
  \bibinfo{pages}{104526} (\bibinfo{year}{2010}).

\bibitem[{\citenamefont{Lebedev and L'vov}(2010)}]{Lebedev2010}
\bibinfo{author}{\bibfnamefont{V.}~\bibnamefont{Lebedev}} \bibnamefont{and}
  \bibinfo{author}{\bibfnamefont{V.}~\bibnamefont{L'vov}},
  \bibinfo{journal}{Journal of Low Temperature Physics}
  \textbf{\bibinfo{volume}{161}}, \bibinfo{pages}{548} (\bibinfo{year}{2010}).

\bibitem[{\citenamefont{Nazarenko}(2013)}]{Nazarenko2013}
\bibinfo{author}{\bibfnamefont{S.}~\bibnamefont{Nazarenko}}
  (\bibinfo{year}{2013}), Private communication.

\bibitem[{\citenamefont{Bradley et~al.}(2006)\citenamefont{Bradley, Clubb,
  Fisher, Gu\'enault, Haley, Matthews, Pickett, Tsepelin, and
  Zaki}}]{Bradley2006}
\bibinfo{author}{\bibfnamefont{D.~I.} \bibnamefont{Bradley}},
  \bibinfo{author}{\bibfnamefont{D.~O.} \bibnamefont{Clubb}},
  \bibinfo{author}{\bibfnamefont{S.~N.} \bibnamefont{Fisher}},
  \bibinfo{author}{\bibfnamefont{A.~M.} \bibnamefont{Gu\'enault}},
  \bibinfo{author}{\bibfnamefont{R.~P.} \bibnamefont{Haley}},
  \bibinfo{author}{\bibfnamefont{C.~J.} \bibnamefont{Matthews}},
  \bibinfo{author}{\bibfnamefont{G.~R.} \bibnamefont{Pickett}},
  \bibinfo{author}{\bibfnamefont{V.}~\bibnamefont{Tsepelin}}, \bibnamefont{and}
  \bibinfo{author}{\bibfnamefont{K.}~\bibnamefont{Zaki}},
  \bibinfo{journal}{Phys. Rev. Lett.} \textbf{\bibinfo{volume}{96}},
  \bibinfo{pages}{035301} (\bibinfo{year}{2006}).

\bibitem[{\citenamefont{Kondaurova and Nemirovskii}(2012)}]{Kondaurova2012}
\bibinfo{author}{\bibfnamefont{L.}~\bibnamefont{Kondaurova}} \bibnamefont{and}
  \bibinfo{author}{\bibfnamefont{S.~K.} \bibnamefont{Nemirovskii}},
  \bibinfo{journal}{Phys. Rev. B} \textbf{\bibinfo{volume}{86}},
  \bibinfo{pages}{134506} (\bibinfo{year}{2012}).

\bibitem[{\citenamefont{Nemirovskii}(2010)}]{Nemirovskii2010}
\bibinfo{author}{\bibfnamefont{S.~K.} \bibnamefont{Nemirovskii}},
  \bibinfo{journal}{Phys. Rev. B} \textbf{\bibinfo{volume}{81}},
  \bibinfo{pages}{064512} (\bibinfo{year}{2010}).

\end{thebibliography}

\end{document}